\documentclass[a4paper,11pt]{article}
\usepackage{jcappub} %,amsmath,amssymb,amsfonts}
\newcommand*\Laplace{\mathop{}\!\mathbin\bigtriangleup}
\usepackage{hyperref}

\newcommand{\rojo}[1]{{#1}}
\newcommand{\be}{\begin{eqnarray}}
\newcommand{\ee}{\end{eqnarray}}
\newcommand{\bea}{\begin{eqnarray}}
\newcommand{\eea}{\end{eqnarray}}

\def\ba{\begin{eqnarray}}
\def\ea{\end{eqnarray}}
\def\be{\begin{equation}}
\def\ee{\end{equation}}

\begin{document}
\title{Constraining gravity theories with the gravitational stability mass}

\author{Camilo Santa V\'{e}lez$^{2,3}$ Antonio Enea Romano$^{1,2,3,4}$ }
\affiliation
{
%${}^2$King's College London, Strand, London, WC2R 2LS, United Kingdom\\
$^1${Theoretical Physics Department, CERN, CH-1211 Geneva 23, Switzerland}\\
${}^2$Instituto de Fisica, Universidad de Antioquia, A.A.1226, Medellin, Colombia\\
${}^{3}$ICRANet, Piazza della Repubblica 10, I--65122 Pescara, Italy \\
${}^{4}$Department of Physics \& Astronomy, Bishop's University\\
2600 College Street, Sherbrooke, Qu\'ebec, Canada J1M~1Z7
% ${}^1$Yukawa Institute for Theoretical Physics, Kyoto University, Kyoto 606-8502, Japan \\

}

%\emailAdd{camilo.santav@udea.edu.co }
\emailAdd{antonio.enea.romano@cern.ch }

% \author{Camilo Santa V\'{e}lez$^{1}$ Antonio Enea Romano$^{1}$ }
% \affiliation
% {
% %${}^1$Yukawa Institute for Theoretical Physics, Kyoto University, Kyoto 606-8502, Japan \\
% ${}^2$Instituto de Fisica, Universidad de Antioquia, A.A.1226, Medellin, Colombia\\
% }

% general result for turn around, check it is equivalent to a point mass =volume integral of \delta\rho
% add calculation of metric of rotational curves in CPT
% integrate t00 and check gives th expected dark matter profile

\abstract{The measurement of the  size of gravitationally bounded structures is an important test of  gravity theories. For a given radius different theories can in fact predict a different gravitational stability mass (GSM) necessary to ensure the stability of the structure in presence of  dark energy. 
We compute the GSM  of gravitationally bounded structures as a function of the radius for different scalar-tensor theories, including $f(R)$ and generalized Brans-Dicke, and compare the theoretical predictions to observational data.
Since the GSM only gives a lower bound, the most stringent  constraints come few objects with a mass lower that the one expected in general relativity. 

The analysis of different observational data sets shows that modified gravity theories (MGT) are compatible with observational data, and in some cases  fit the data better than general relativity (GR), but the latter is not in strong tension with the observations. The data presently available does not provide a statistically significant evidence of the need of a modification of GR, with the largest deviation of order $2.6 \,\sigma$ for the galaxy cluster NGC5353/4. Due to the limited number of objects not satisfying the GR bound, for these structures it may be important to take into account non gravitational physics or deviations from spherical symmetry.
}

 %, checking it is consistent with the result obtained in static coordinates for the SDS metric and for other SSS metrics.  

%\end{abstract}
\maketitle

\section{Introduction}
Modified gravity theories (MGT) have been extensively investigated as possible solutions of some of the unsolved puzzles of the observed Universe, such as the nature of dark energy or dark matter \cite{DeFelice:2010aj, Starobinsky:2007hu, Hu:2007nk, Nicolis:2008in, Burrage:2016yjm, Clifton:2011jh, Frusciante:2018aew, Kase:2018nwt}, and can also provide models of cosmic inflation  \cite{Starobinsky:1980te, Tsujikawa:2014mba, Ohashi:2012wf, DeFelice:2011jm, Bamba:2015uma} in very good agreement with observations \cite{Akrami:2018odb}. It is thereof important to set constraints on MGT using different types of observations, and one important test is provided by the stability of cosmic structure \cite{Bhattacharya:2016vur}, in particular the turn around radius, i.e. the maximum size of a spherically symmetric gravitationally bounded object in presence of dark energy. The effects of the modification of gravity were considered previously in the case of the Brans-Dicke theory \cite{Bhattacharya:2016vur,Bhattacharya:2015iha} and some classes of Galileion theories \cite{Bhattacharya:2015chc}, while here  we consider a wider class of scalar tensor-theories, including among  others $f(R)$, Generalized Brans-Dicke and quintessence theories.
For convenience in the comparison with observational data, in this paper we compute a closely related quantity, the \textit{gravitational stability mass} (GSM) necessary to ensure the stability of a gravitationally bounded structure of given radius.

The first calculations of the turn around radius were based on the use of static coordinates \cite{Bhattacharya:2015iha, Pavlidou:2013zha}, which can be related to cosmological perturbation with respect to the Friedmann metric via an appropriate background coordinate transformation, allowing to establish a gauge invariant definition of the turn around radius \cite{Velez:2016shi}. 
Here we show that the use of cosmological perturbations theory is more convenient and allows to find general theoretical predictions which can be applied to a wide class of scalar-tensor theory.

The theoretical predictions are compared to different set of observations, finding that MGT are in better agreement with observations for few data points, but without a strong evidence of a tension with GR, with the largest deviation of $\approx2.6 \,\sigma$ for the galaxy cluster NGC5353/4.

%Note that the gravitational stability mass only allows to cons parameters defining the different theories, since the main constraints come from the objects with the lowest observed mass for a given observed radius. 
 
%\section{Observational constraints on the minimal mass}

\section{Effective gravitational constant in scalar-tensor theories}
%iven the broad spectrum of modified gravity theories, we are only going to focus in a subset of Scalar-Tensor theories which couple ascalar field non-minimally with the Ricci scalar. 

We will focus on the class of scalar-tensor theories defined by the  action \cite{0705.1032}:
\begin{equation}\label{accion}
    \mathcal{S}=\int d^4x \sqrt{-g}\left[\frac{1}{2}f(R,\phi,X)-2\Lambda+\mathcal{L}_m\right],
\end{equation}
where $\Lambda$ is the bare cosmological constant, $\phi$ is a scalar field and $X=-\frac{1}{2}\partial_\mu\phi \partial^\mu \phi$ is the scalar field's kinetic term, and we used a system of units in which $c=1$. %The gravitational field equations obtained by taking the variation of the above action with respect to the metric are.
%\bea
%\eea

For non-relativistic matter with energy-momentum tensor 
\be
\delta T_0^0=\delta\rho_m, \,\, \delta T_i^0=-\rho_m v_{m,i},
\ee
where $v_m$ is the matter velocity potential, and using the metric for scalar perturbations in the Newton gauge 
\begin{equation}\label{Newton}
    ds^2=-(1+2\Psi)dt^2+a^2(1-2\Phi)\delta_{ij}dx^i dx^j,
\end{equation}
the Fourier's transform of the  Einstein's equations give the modified Poisson  equation \cite{0705.1032} 
\begin{equation}\label{Psi}
    \Psi_k=-4\pi \tilde{G}_{eff}\frac{a^2}{k^2}\rho_m\delta_k, %\lable{Poiss}
\end{equation}
where $k$ is the comoving wave number, the subscript $k$ denotes the corresponding Fourier's modes, and $\delta_k$ %=\frac{\delta \rho_k}{\rho_m}+3Hav_m$ 
is the gauge-invariant matter density contrast. The quantity $\tilde{G}_{eff}$, normally interpreted as the effective gravitational ``constant", is given by \cite{0705.1032}
\begin{equation}\label{Geff}
    \tilde{G}_{eff}=\frac{1}{8\pi  F}\frac{f_{,X}+4\left(f_{,X}\frac{k^2}{a^2}\frac{F_{,R}}{F}+\frac{F_{,\phi}^2}{F}\right)}{f_{,X}+3\left(f_{,X}\frac{k^2}{a^2}\frac{F_{,R}}{F}+\frac{F_{,\phi}^2}{F}\right)},
\end{equation}
where $F=\frac{\partial f}{\partial R}$.

%The other gauge-invariant potential $\Phi$ can be obtained from  the  anisotropy  parameter:
%\begin{equation}\label{eta}
%    \eta=\frac{\Psi-\Phi}{\Phi}=\frac{2f_{,X}\frac{k^2}{a^2}\frac{F_{,R}}{F}+2\frac{F_{,\phi}^2}{F}}{f_{,X}+2f_{,X}\frac{k^2}{a^2}\frac{F_{,R}}{F}+2\frac{F_{,\phi}^2}{F}},
%\end{equation}
%which corresponds to the postnewtonian parameter
%\begin{equation}
%    \gamma=\frac{1}{1+\eta}.
%\end{equation}
%Equations (\ref{Psi}-\ref{eta}) completely determine the metric \eqref{Newton} which can then be used to compute observables. %We will focus  turn around radius establishes a limit in the size of cosmological structures, given by the radius at which the effect of the expanding universe balances with the gravitational collapse.
\section{Gravitational stability mass}
According to \cite{Velez:2016shi} and \cite{Faraoni:2015zqa} the turn around radius  can be computed from the gauge invariant Bardeen's potentials by solving the equation
\begin{equation}
    \ddot{a}r-\frac{\Psi^\prime}{a}=0,\label{rtar}
\end{equation}
where the dot and the prime denote derivatives respectively respect to time and the radial coordinate.
Note that the above condition is independent of the gravity theory since it is only based on the use of the metric of cosmological perturbations in the Newton gauge, and  \textit{it does not} assume any gravitational field equation. We can take advantage of the generality of eq.(\ref{rtar}) and  apply it to any gravity theory, in particular to the theories defined in eq.(\ref{accion}).

% The effective gravitational constant defined in eq.(\ref{Geff}) is in momentum space, but to first order in perturbations, only the momentum independent part survives, since on the r.h.s. of eq.(\ref{Psi}) $\delta_k$ is already a first order quantity, i.e. to first order in perturbations we only need

% \begin{equation}\label{Geff}
%     G_{eff}=\frac{1}{8\pi  F}\frac{f_{,X}+4\left(\frac{F_{,\phi}^2}{F}\right)}{f_{,X}+3\left(\frac{F_{,\phi}^2}{F}\right)} \,.
% \end{equation}

For the theories we will consider and in the sub-horizon limit,  we can then take the inverse Fourier's transform of eq.(\ref{Psi})  to get a real space  modified Poisson's equation of the form

\be
\Laplace \Psi =- 4\pi G_{eff} \rho_m  \delta \,. \label{PoissR}
\ee

The gravitational potential outside a spherically symmetric object of mass $m$ is then obtained by integrating the modified Poisson's equation (\ref{PoissR}) 
\begin{equation}\label{PsiSol}
    \Psi=-\frac{G_{eff} m}{r},
\end{equation}
which substituted in  eq.\eqref{rtar}  allows to derive a general expression for the turn around radius for all the scalar-tensor theories defined  in eq.(\ref{accion})  
\begin{equation}\label{TAR}
    r_{TAR}=\sqrt[3]{\frac{3 G_{eff}m}{\Lambda}} \,. 
\end{equation}
It is convenient to define the  ratio between the Newton constant $G$ and the effective gravitational constant as $\Delta=G/G_{eff}$ and the gravitational stability mass (GSM) as:
\begin{equation}
   m_{gs}=\frac{\Lambda r_{obs}^3}{3G_{eff}}=m_{GR}\Delta \,,
\end{equation}
where $m_{GR}(r_{obs})=\Lambda r_{obs}^3/3 G$ is the value of the GSM predicted by GR.
Any  object of  mass $m_{obs}$ should  have a radius $r_{obs}<r_{TAR}(m_{obs})$, or viceversa any gravitational bounded object of radius $r_{obs}$ should have a mass larger than $m_{gs}$
\be
m_{obs}(r_{obs})>m_{gs}(r_{obs})=\frac{\Lambda r_{obs}^3}{3 G_{eff}}=m_{GR}(r_{obs})\Delta\, \label{mgsm}.
\ee
 
In fact objects of size $r_{obs}$ with a mass smaller than $m_{gs}(r_{obs})$ would not be gravitationally stable, since the effective force due to dark energy will dominate the attractive gravitational force. 
%According to current cosmological observations given by \cite{1310.1920} the turn around radius must be at least 90\% the one predicted by general relativity. 
%\begin{equation}
%    \sqrt[3]{\frac{3G_{eff} m}{\Lambda}}\geq 0.9 \sqrt[3]{\frac{3G_0 m}{\Lambda}}
%\end{equation}
%\begin{equation}\label{Limit}
%    \delta\gtrapprox 0.729
%\end{equation}

In order to compare theories to experiment, it is important to establish what is the size of gravitationally bounded structures, and for  this purpose  the caustic method has been developed \cite{Yu:2015dqa}, showing good accuracy when applied to simulated data. In the rest of this paper we will use the results of the application of this method to set constraints on the parameters of the different MGT. 

Galaxy clusters data \cite{Lee,Rines:2012np} can be used to set upper bounds on GSM, and  to consequently set constraints on $G_{eff}$, since from eq.(\ref{mgsm}) we get %$\Delta<m_{obs}/m_{GR} .$
\be
\Delta<\frac{m_{obs}}{m_{GR}}\, .
\ee
%\begin{center}
% \begin{tabular}{||c c c||} 
% \hline
% Outliers & $\delta$ & $\sigma_\delta$ \\ [0.5ex] 
% \hline\hline
% Included & 0.71695942 & 0.00734177\\ 
% \hline
% Excluded & 1.19407806 & 0.06654307 \\ [1ex] 
% \hline
%\end{tabular}
%\end{center}
\section{Gravity theory independent constraints}
\begin{figure}[h]
    \includegraphics[scale=0.37]{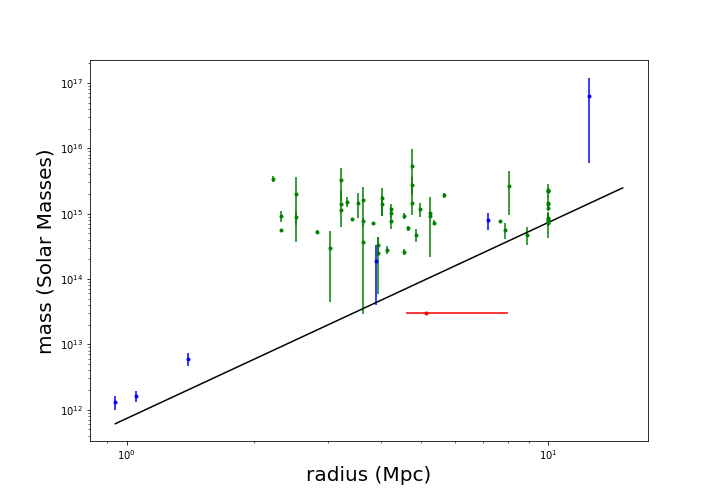}
    \caption{Observed masses and radii of  galaxy clusters are compared to the GR prediction (black line). Vertical green lines represent the errors on the estimation of the masses from \cite{Rines:2012np} and blue lines correspond to other cosmological structures in \cite{Pavlidou:2013zha}. The object with the most significant  deviation is NGC5353/4 plotted in red \cite{Lee}, which is shown in more detail in fig.(\ref{fig:my_label2}).}
    \label{fig:my_label}
\end{figure}

\begin{figure}[h]
    \includegraphics[scale=0.3]{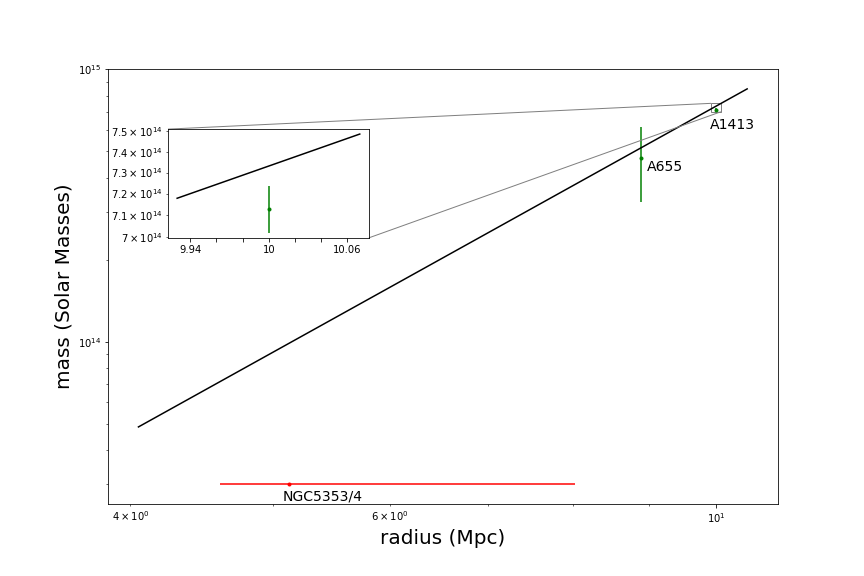}
    \caption{Observed masses and radii of  the A655, A1413 and NGC5353/4 galaxy clusters. These are the objects with the most significant deviation from the GR prediction, respectively of order $0.19\sigma$ for A655, $1.84\sigma$ for A1413 (see inset) and $2.61\,\sigma$ for NGC5353/4. 
    %The blue lines correspond to scalar-tensor theories with a value of $\Delta$ such that $m_{gs}=m_{obs}$ for the each cluster.
    }
    \label{fig:my_label2}
\end{figure}

Before considering the constraints on specific gravity theories in the next sections,  we can derive some general gravity theory independent constraints for $G_{eff}$. 
Note that since the GSM only gives a lower bound for the mass of an object of given radius, we cannot fit the data points one by one, since  each different gravity theory predicts a range of masses $m>m_{gs}(r_{obs})$, not a single value. Consequently most the constraints come from the objects with the lowest masses.
Most of the data  are consistent with GR, except few data points corresponding to the galaxy clusters A655, A1413 and NGC5353/4, which give respectively $\Delta<0.9162 \pm 0.2812$, $\Delta<0.9723 \pm 0.0151$ and $\Delta<0.0969\substack{+0.3215\\ -0.0178}$, as shown in fig.(\ref{fig:my_label}-\ref{fig:my_label2}). The errors have been obtained by Gaussian propagation from the errors on $m_{obs}$ corresponding to $r_{MAX}$ in \cite{Rines:2012np} for A655 and A1413, and from the probability distribution for the size of NGC5353/4 in \cite{Lee} using the normalization relation 
\begin{equation}
    \int \rho_r(r) dr=\int \rho_\Delta[\Delta(r)]d\Delta=1
\end{equation}
where $\rho_r$ and $\rho_\Delta$ are the probability density functions of the size $r$ and $\Delta$ respectively. The lower and upper bound are taken from the symmetric two-tail limits on the distribution and the main value is taken as the maximum likelihood estimate for $\Delta$.
%**add sigma order for each, a655? 3/4?
The tightest constraints for GR come from A1413 and NGC5353/4, whose deviation from GR is  respectively of order $1.84 \,\sigma$ and  $2.61 \,\sigma$, implying  that there is not a very strong evidence of the need of a modification of GR.

%\rojo{
%This is in agreement with both \cite{Avilez:2013dxa} and \cite{Lin:2010hk} at $3\sigma$, however they don't exclude General Relativity.
%}

% The values of $\Delta$ including the mass uncertainty which are consistent with all the data points are
% %\begin{equation}
% %    \delta=1.028525297656162
% %\end{equation}
% \begin{center}
%  \begin{tabular}{|c c|} 
%  \hline
%  $\Delta$ & 1.0285\\ 
%  \hline
%  $\Delta+\sigma$ & 1.0128\\ 
%  \hline
%  $\Delta+2\sigma$ & 0.9976\\ 
%  \hline
%  $\Delta+3\sigma$ & 0.9829\\ 
%  \hline
% \end{tabular}
% \end{center}
%This value of $\delta$ represents the blue curve in the following graph
%Two other cases are considered, this first fit includes all the data from the catalogue 
%\begin{figure}[h]
%    \includegraphics[scale=0.35]{fitMax}
%    \caption{Observed masses and radii of  galaxy clusters are compared to different gravity theories predictions. The black line is for GR and the blue line corresponds to scalar tensor theories with a value of $\Delta$ consistent with all data points. The zoomed region shows in more details the galaxy clusters A655 and A1413. The latter is the object with the largest difference with respect to the GR prediction, but is still within the $2 \, \sigma$ confidence level. }
%    \label{fig:my_label}
%\end{figure}
%\begin{equation}\label{delta}
%    \delta=0.71695942,
%\end{equation}
%\begin{equation}\label{sigma}
%    \sigma_\delta=0.00734177.
%\end{equation}
%\begin{figure}[h]
%    \includegraphics[scale=0.5]{fitOut}
%    \label{fig:my_label}
%\end{figure}
%The second fit excludes data points in which $\sigma_m/m$ is an outlier defined by the first 11\% percentile rank
%\begin{equation}\label{delta}
%    \delta=1.19407806,
%\end{equation}
%\begin{equation}\label{sigma}
%    \sigma_\delta=0.06654307.
%\end{equation}
%\begin{figure}[h]
%    \includegraphics[scale=0.5]{fit}
%    \caption{The black curve corresponds to the turn-around radius predicted by General Relativity. The blue curve is the best fit and the green curves correspond to 1$\sigma_\delta$ variations. The red datapoints are the outliers which are ignored in this fit.}
%    \label{fig:my_label}
%\end{figure}
%This implies that the effective gravitational constant must be larger than the one given by general relativity. 

%It should also be noted observed structure have not necessarily achieved their maximum size, conservative approximation because it assumes that these cosmological structures have reached their maximum size. If they are allowed to expand even further, the effective gravitational constant must be even larger to accomodate their larger turn around radii. These results can be applied to several specific modified gravity theories such as Brans-Dicke, $f(R)$ and Quintessence.

\section{\texorpdfstring{$f(R)$}{Lg} theories}
In this case the action is independent of the scalar field, and in the Jordan frame is
\begin{equation}
    \mathcal{S}=\int d^4x \sqrt{-g}\left[\frac{1}{2}f(R)-2\Lambda+\mathcal{L}_m\right]\, ,
\end{equation}
with the effective gravitational constant given by
\begin{equation}
    \tilde{G}_{eff}=\frac{1}{ 8\pi F}\frac{1+4\frac{k^2}{a^2}\frac{F_{,R}}{F}}{1+3\frac{k^2}{a^2}\frac{F_{,R}}{F}}\, .
\end{equation}
On sub-horizon scales ($\frac{k^2}{a^2}\frac{F_{,R}}{F}\gg 1$) it reduces  to \cite{0705.1032}
\begin{equation}
    \tilde{G}_{eff}=G_{eff}=\frac{1}{6\pi F}\, ,
\end{equation}
%\begin{equation}
%    \Delta=\frac{3F}{4}\, ,
%\end{equation}
and the turn around radius is given by
\begin{equation}
    r_{TAR}=\sqrt[3]{\frac{ m}{2\pi \Lambda F}}\, ,
\end{equation}
which corresponds to  this expression for the GSM
\begin{equation}
m_{gs}=2\pi\Lambda F r_{obs}^3\, .
\end{equation}
\rojo{Observational data imply $F<(0.0486\pm 0.0149)G^{-1}$ for A655, $F<(0.0516\pm 0.0008)G^{-1}$ for A1413, and $F<0.0051\substack{+0.0009\\ -0.0171}$  for  NGC5353/4. It can noted that GR is not incompatible with observations, since the tightest constraint on $F$, corresponding to NGC5353/4 is $2.61 \,\sigma$ away from the GR limit  $F=(6\pi G)^{-1}\approx 0.0531 G^{-1}$.}
%\begin{center}
% \begin{tabular}{||c c c||} 
% \hline
% Outliers & F & $\sigma_F$ \\ [0.5ex] 
% \hline\hline
% Included & 0.955946 & 0.00978903\\ 
% \hline
% Excluded & 1.5921 & 0.0887241 \\ [1ex] 
% \hline
%\end{tabular}
%\end{center}
%From \eqref{Limit}
%\begin{equation}
%    F\lessapprox 1.82899
%\end{equation}

\section{\texorpdfstring{$R^n$}{Lg} theories}
For these theories the action is given by
\begin{equation}\label{actfR}
    f(R,\phi,X)=\frac{1}{8\pi G}R+\frac{\alpha}{8\pi G} R^n,
\end{equation}
and the corresponding  effective gravitational constant is
\begin{equation}
    G_{eff}=\frac{4 G}{3 (1+ n \alpha R^{n-1})}=\frac{4 G}{3(1+\alpha \beta)},
\end{equation}
where $\beta=n R^{n-1}$,
which gives the following expressions for the turn around radius and GSM
\begin{equation}\label{tarfR}
    r_{TAR}=\sqrt[3]{\frac{4 G m}{\Lambda [1+  \alpha \beta]}}\, ,
\end{equation}
\begin{equation}\label{gsmfR}
    m_{gs}=\frac{\Lambda r_{obs}^3 [1+\alpha \beta]}{4 G}\, .
\end{equation}

In  fig.(\ref{fig:fitFr}) we plot the regions of the $(\alpha$,$\beta)$ parameters space satisfying the condition $m_{obs}>m_{gs}$, with the strongest constraints coming from the NGC5353/4 galaxy cluster. 
\begin{figure}[h]
    \includegraphics[scale=0.42]{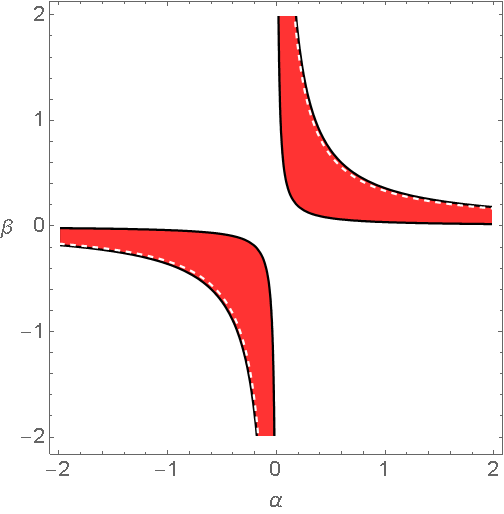}
    \caption{Allowed regions of the ($\alpha,\beta$) parameters space for $R^n$ theories, respectively in units of $Mpc^{2n-2}$ and $Mpc^{2-2n}$ according to \cite{Avilez:2013dxa}. The white dashed line corresponds to GR.}
    \label{fig:fitFrSkordis}
\end{figure}
\begin{figure}[h]
    \includegraphics[scale=0.42]{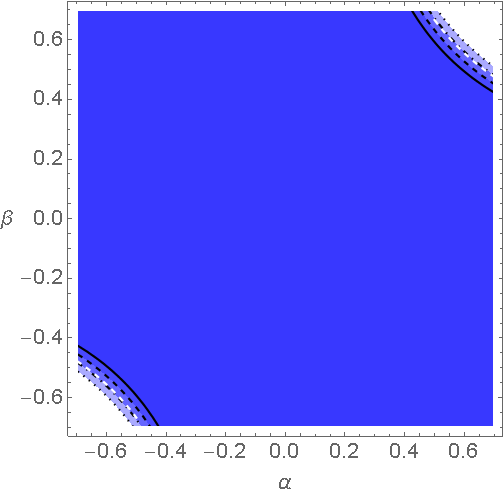}
    \caption{Allowed regions of the ($\alpha,\beta$) parameters space for $R^n$ theories, respectively in units of $Mpc^{2n-2}$ and $Mpc^{2-2n}$. This constrains come from the galaxy cluster A1413. The dark blue region corresponds to $m_{obs}>m_{gs}$ and the other colours to three different confidence bands defined by $m_{obs}+n\,\sigma_m>m_{gs}$, delimited by dashed (n=1), dot-dashed (n=2), and dotted (n=3) lines respectively.
    The continuous black line corresponds to the parameters which give $m_{obs}=m_{gs}$. The white dashed line corresponds to GR.}
    \label{fig:fitFrVasi}
\end{figure}
\begin{figure}[h]
    \includegraphics[scale=0.42]{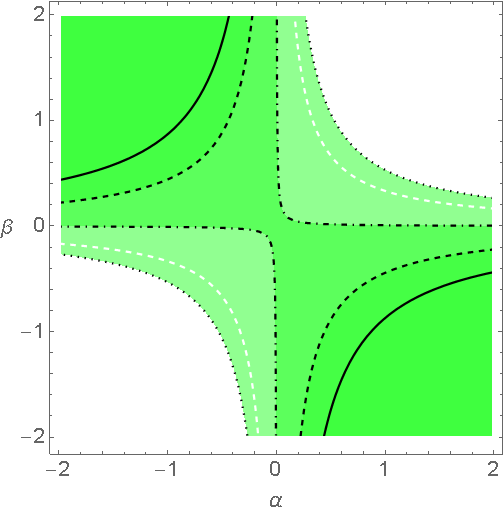}
    \caption{Allowed regions of the ($\alpha,\beta$) parameters space for $R^n$ theories, respectively in units of $Mpc^{2n-2}$ and $Mpc^{2-2n}$. The main constrains come from the galaxy cluster NGC5353/4 studied in  \cite{Lee}. The dark green region corresponds to $m_{obs}>m_{gs}$,  and other regions are the confidence bands as defined in fig.(\ref{fig:fitFrVasi}). The continuous black line corresponds to the parameters which give $m_{obs}=m_{gs}$. The white dashed line corresponds to GR.}
    \label{fig:fitFrLee}
\end{figure}
\begin{figure}[h]
    \includegraphics[scale=0.42]{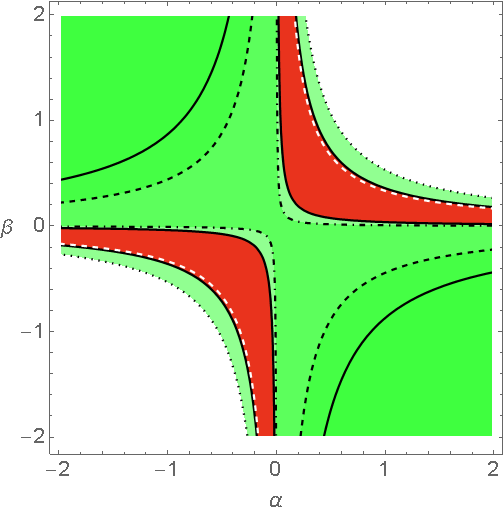}
    \caption{Allowed regions of the ($\alpha,\beta$) parameters space for $R^n$ theories, respectively in units of $Mpc^{2n-2}$ and $Mpc^{2-2n}$. This figure shows the overlapping regions between fig.(\ref{fig:fitFrSkordis}) and fig.(\ref{fig:fitFrLee}).}
    \label{fig:fitFr}
\end{figure}

\section{Generalized Brans-Dicke}

These theories \cite{DeFelice:2010jn,Roy:2017mnz} are  a generalization of Brans-Dicke theory \cite{Brans:1961sx}, with a more general kinetic term, defined by the action 
\begin{equation}
    f(R,\phi,X)=\frac{\phi}{8\pi G} R+\frac{g(\phi)}{4\pi G} X\, .
\end{equation}
After decomposing the scalar field as the sum of a homogeneous background component and a space dependent perturbative part according to
\begin{equation}
\phi(t,x)=\overline{\phi}(t)+\delta\phi(t,x) \,,
\end{equation}
at leading order in perturbations the effective gravitational constant is given by
\begin{equation}
G_{eff}=\frac{4+2\phi_0 \, g_0}{3\phi_0 +2\phi_0^2 \, g_0}G \, ,
\end{equation}
% \begin{equation}
% \Delta=\frac{3\phi_0 +\phi_0^2 \,g_0}{4+\phi_0 g_0}
% \end{equation}
where $\phi_0=\bar{\phi}(t_0)$, $g_0=g(\phi_0)$, and $t_0$ is the cosmic time corresponding to the red-shift of the observed structure. 

\begin{figure}[h]
    \includegraphics[scale=0.42]{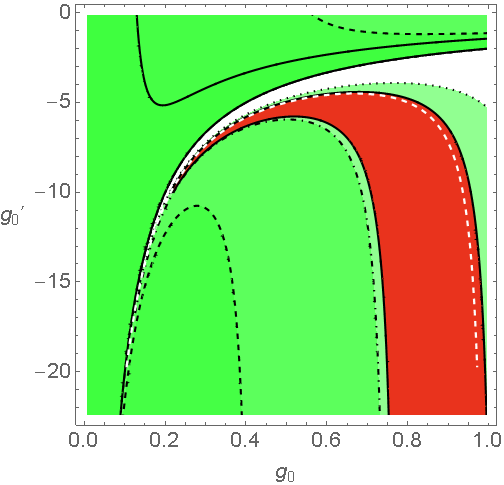}
    \caption{Allowed regions of the $(\phi_0,g_0)$ dimensionless parameters space, for generalized Brans-Dicke theories. The main constrains come from the galaxy cluster NGC5353/4 studied in \cite{Lee}. The dark green region corresponds to $m_{obs}>m_{gs}$,  and the other regions are the confidence bands as defined in fig.(\ref{fig:fitFrVasi}). The continuous black line corresponds to the parameters which give $m_{obs}=m_{gs}$. The white dashed line corresponds to the GR limit.}
    \label{fig:fitGenBrans}
\end{figure}

The corresponding turn around radius and GSM are given by
\begin{equation}
    r_{TAR}=\sqrt[3]{\frac{3Gm}{\Lambda}\frac{4+2\phi_0 g_0}{3\phi_0 +2\phi_0^2 g_0}}\, ,
\end{equation}
\begin{equation}
    m_{gs}=\frac{r_{obs}^3\Lambda}{3G}\frac{3\phi_0 +2\phi_0^2 g_0}{4+2\phi_0 g_0}\, .
\end{equation}
The regions of the $(\phi_0$, $g_0)$ parameters space satisfying the condition $m_{obs}>m_{gs}$ are shown in fig.\ref{fig:fitGenBrans} for the strongest constraints, which come from the NGC5353/4 galaxy cluster.

\section{Quintessence}
\begin{figure}[h]
    \includegraphics[scale=0.42]{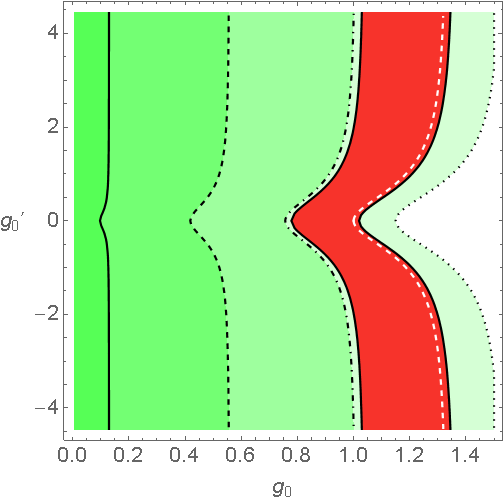}
    \caption{Allowed regions of the ($g_0,g_0'$) dimensionless parameters space for Quintessence. The main constrains come from the galaxy cluster NGC5353/4 studied in \cite{Lee}. The dark green region corresponds to $m_{obs}>m_{gs}$,  and the other regions are the confidence bands as defined in fig.(\ref{fig:fitFrVasi}). The continuous black line corresponds to the parameters which give $m_{obs}=m_{gs}$. The white dashed line corresponds to the GR limit.}
    \label{fig:fitQuint}
\end{figure}

The action of Quintessence is given by
\begin{equation}
    f(R,\phi,X)=\frac{g(\phi)}{8\pi G}R-\frac{1}{4\pi G}X\, ,
\end{equation}
and in this case the effective gravitational constant is
\begin{equation}
    G_{eff}=\frac{G}{g_0}\frac{2g_0+4g_0'^2}{ 2g_0+3 g_0'^2 }\, ,
\end{equation}

where $g_0=g(\phi_0)$, $g_0'=g'(\phi_0)$, $\phi_0=\overline{\phi}(t_0)$ and the 
turn around radius and GSM are given by
\begin{equation}
    r_{TAR}=\sqrt[3]{\frac{3 G m}{g_0\Lambda}\frac{2g_0+4g_0'^2}{ 2g_0+3  g_0'^2 }}\, ,
\end{equation}
\begin{equation}
    m_{gs}=\frac{g_0\Lambda r_{obs}^3}{3G}\frac{ 2g_0+3 g_0'^2 }{2g_0+4g_0'^2}\, .
\end{equation}
We plot in  fig.(\ref{fig:fitQuint}) the regions of the $(\phi_0$, $g_0)$ parameters space satisfying the condition $m_{obs}>m_{gs}$, with the strongest constraints coming from the NGC5353/4 galaxy cluster. 
%\section{Related Work}
%In \cite{DeFelice:2010aj} $f(R)$ theories with light Scalaron masses are shown to be incompatible with solar system tests. Therefore, it's necessary to introduce the chameleon field to mask the modification of gravity at local scales.On the other hand, the authors of \cite{Song:2006ej} study the large scale structure of $f(R)$ gravity and, depending on the sign of $f_{RR}$, the theory might agree with observations of the CMB. If $f_{RR}$ is negative there is no agreement with CMB observations, if it's positive the low observed quadrupole moment can be explained and $f_{RR}=0$ is the $\Lambda CDM$ limit.
%In \cite{Chiba:2003ir} the authors constrain both Brans-Dicke and $f(R)$ gravity, specifically $1/R$ and Starobinsky's $R^2$ models. A lower bound is found for the Brans Dicke parameter $\omega_{BD}>3500$

\section{Conclusions}
 We have derived the theoretical prediction of the gravitational stability mass for a wide class of scalar-tensor theories including $f(R)$ and generalized Brans-Dicke. Most of observations are consistent with GR except the galaxy clusters A655, A1413 and NGC5353/4, which have masses smaller than the GR prediction.
 %The deviation from GR is of order  A1413 the difference between the estimated mass and the GR prediction is of order $2\sigma$, and we have computed the values of the parameters of different MGT which could provide a better agreement between theoretical prediction and observations for this object. 
 The tightest constraints for GR come from A1413 and NGC5353/4, whose deviation from GR is  respectively of order $1.84\,\sigma$ and  $2.61\,\sigma$, implying  that there is  no statistically significant evidence of the need of a modification of GR.
 In the future it will be important to increase the size of the data sets used for testing different gravity theories, including for example the observations of the upcoming Euclid \cite{Laureijs:2011gra,Tereno:2015hja,Scaramella:2015rra,Amiaux:2012bt} mission. 
 
 In this paper we have assumed a fixed value of the cosmological constant,   but in the future it could be interesting to investigate the interplay between MGT and dark energy by fitting both at the same time, and assess the existence of a possible degeneracy between the two.
 Due to the limited number of objects not satisfying the GR bound, for these structures it may be important to take into account non gravitational physics or deviations form spherical symmetry \cite{Giusti:2019uez,PCP}.
 
 \section{Acknowledgements}
 We thank Vasiliki Pavlidou for interesting discussions. This work was supported by the Sostenibilidad program of UDEA.

\bibliography{Bibliography}
\bibliographystyle{h-physrev4}

\end{document}